\documentclass[12pt]{article}
\newcommand{\beq}{\begin{equation}}
\newcommand{\eeq}{\end{equation}}
\newcommand{\bra}{\begin{array}}
\newcommand{\era}{\end{array}}

\newcommand{\te}{\theta}
\newcommand{\al}{\alpha}
\newcommand{\ga}{\gamma}
\newcommand{\de}{\delta}

\newcommand{\Om}{\Omega}

\newcommand{\ep}{\epsilon}
\usepackage{latexsym}
\author{Jamila Douari\footnote{jdouari@excite.com}\\ \\
\small\it Stellenbosch Institute for Advanced Study, Private Bag X1,\rm\\
\small\it Matieland, Stellenbosch, 7601, South Africa\rm }
\title{Exotic Particles and Generalized Maxwell theory on Fuzzy Two-Sphere}

\begin{document}
\maketitle
%\vspace*{0.5cm}
%PACS:
\vskip1cm
Keywords: Generalized connection, Exotic particles, Fuzzy Two-Sphere.
\vspace*{1cm}
\section*{Abstract}
\hspace{.3in}We consider generalized Maxwell theory and spherical
D2-brane. The model is built by introducing a generalized connection
put at the origin of two-sphere to describe anyons instead of
Chern-Simons term. The energy obtained in this model is very special
since the gauge field is dynamic and its energy dominates when the
radius of fuzzy two-sphere goes to infinity or if we take large
number of charges. Consequently, D2-brane gets high energy. The
static potential for two opposite charged exotic particles described
by generalized Maxwell theory is found to have screening nature on
fuzzy two-sphere instead of confinement which is a special property
of the system on the plane.
\section{Introduction}
\hspace{.3in}Various brane configurations have attracted much
attention over the recent years and several papers have been devoted
to the study of a relationship between noncommutative geometry
\cite{ncg} and string theory \cite{ncst} and the relationship
between D-branes with different dimensions as well
\cite{brane,fuzzyQHE}. The appearance of noncommutative geometry in
string theory can be understood from a different point of view. For
example, a D2-brane can be constructed from multiple D0-branes by
imposing a noncommutative relation on their coordinates in matrix
theory or under the strong magnetic field the world volume
coordinates of a D2-brane become noncommutative by considering the
quantum Hall system \cite{fuzzyQHE} and the magnetic field charge is
interpreted as the number of D0-branes.

In this work, we consider exotic particles described by generalized
Maxwell theory in which we introduce a generalized connection in
fuzzy two-sphere which has a dual description in terms of an abelian
gauge field on a spherical D2-brane and is interpreted as a bound
state of a spherical D2-brane and D0-branes. The exotic particles
are known as excitations and quasi-particles or anyons; i.e.
fermions (bosons) carrying odd (even) number of elementary magnetic
flux quanta \cite{quanta}. They are living in two-dimensional space
as composite particles having arbitrary spin, and they are
characterized by fractional statistics which are interpolating
between bosonic statistics and fermionic one \cite{quanta,anyon}.
One of the field-theories describing anyons is the model, where the
matter is interacting with the Chern-Simons (CS) gauge field
\cite{GCS}. In the reference \cite{GConn}, Stern has introduced
another approach to treat anyons that does not require the CS term,
but introduces a generalized connection with which the conserved
U(1) current is coupled in a gauge invariant way \cite{Any}. In this
model the gauge field is dynamic and the potential has the confining
nature which makes the model different \cite{GC}.

This paper is devoted to treat the same system but on two-sphere.
Among the main results in this work is the change of the potential's
nature; there is no confinement nature any more, and the
disappearance of the confinement in two-sphere case for the exotic
system is very interesting result. It was shown in \cite{conf} that
compact Maxwell theory in (2+1)-dimensions confines permanently
electric test charges and the usual two-dimensional Coulomb
potential is $V (R) \sim lnR$. Since the electrostatic potential has
the form $V (R)\sim R$ and holds for all values of the gauge
coupling, the compact (2+1)-dimensional Maxwell theory does not
exhibit any phase transition, i.e., the confinement is permanent. In
the present paper, the things are changed by treating the exotic
system in high dimensions and $V(R)\sim \frac{1}{R}$ with $R$ is the
distance between two opposite charged exotic particles. Another
important result we get is at the level of energy; D2-brane gets
high energy if the radius $r$ of the two-sphere goes to infinity and
it is higher if the number of charges is large which makes the
system very special.

The main results of this work: The fuzzy two-sphere is realized as
one of D2-brane descriptions with special properties because of the
generalized Maxwell theory. Among these properties we get the energy
of gauge field dominates when the radius of fuzzy two-sphere goes to
infinity, then the energy of flat D2-brane which is a dual of fuzzy
two-sphere becomes high which is different from the case of quantum
Hall effect (QHE) where the energy of flat D2-brane goes to zero. An
important remark is that our system could be identified to QHE in
high dimensions only if the radius $r$ and the number of charges $N$
go to zero. Also, what makes the model very different and very
special is that the potential loses the confining nature in fuzzy
two-sphere case.
\section{Generalized Connection and Anyons}
\hspace{.3in}The simplest way to realize fractional statistics
characterizing anyons in three-dimensional space-time is usually by
adding a Chern-Simons term to the action. Recently, a novel way was
introduced in \cite{GConn} to describe anyons without a Chern-Simons
term. Thus, a generalized connection was considered in
(2+1)-dimensions denoted $A_\mu^\te$, $\mu=0,1,2$. The gauge theory
is defined by the following Lagrangian \beq
L_\te=-\frac{1}{4}F_{\mu\nu}F^{\mu\nu}+J^\mu A_\mu^\te \eeq with
$A_\mu^\te\equiv A_\mu +\frac{\te}{2}\ep_{\mu\nu\rho}F^{\nu\rho}$
and $\te$ is real parameter in Minkowski space. The Lagrangian
$L_\te$ describes Maxwell theory that couples to the current via the
generalized connection rather than the usual one. This coupling is
gauge invariant as long as $J^{\mu}$ is a conserved external
current. In this theory, the gauge fields are dynamic and the
canonical moments are $\pi^\mu =F^{\mu 0}+\te\ep^{0\mu\nu}J_\nu$
which results in the usual primary constraint $\pi^0 =0$ and $\pi^i
=F^{i 0}+\te\ep^{0ij}J_j$ ($i,j=1,2$). Thus the magnetic field is
$B=\ep_{ij}\partial^i A^j$ and the electric field is $E^i =\pi^i
-\te\ep^{ij}J_j$.

Now, accordingly to (1), the equations of motion for $A_\mu$ give
\beq
\partial^\nu \partial_\nu A_\mu =J_\mu+ \te\ep_{\mu\nu\rho }\partial^\nu J^\rho.
\eeq Then, we consider the simplest case of a static pointlike
particle located at the origin which is described by $J^0
=e\de^{(2)}(x)$. By solving (2) for the gauge field one finds
$$
A_0 =\frac{ln r}{2\pi}, \phantom{~~~~~}A_1 =\frac{\te x_2}{2\pi
r^2}, \phantom{~~~~~}A_2 =\frac{\te x_1}{2\pi r^2},$$ with $r^2
=x_1^2 + x_2^2$. This background describes one unit of an
electrically charged particle and an infinitely thin magnetic flux
with total flux $\te$ both located at the origin and the shift in
the statistics of the particle is fixed by the Aharonov-Bohm effect
to be \beq \Delta\phi =\te. \eeq We note that in the case of
Chern-Simons theory, the phase is two times $\te$ and this is due to
the fact that the charged particle is winding around a magnetic flux
while in the present theory we also have the contribution of a flux
tube winding around the charged particle. Another reason is that
with $A_\mu^\te$ construction a long range electric field is also
generated which couples to the current and gives exactly the same
phase.

For a static charged particle located at the origin and $J^i =0$,
the static electromagnetic fields are \beq \bra{ll}
B(x)=e\te\de^{(2)}(x)\\
E_i (x)=-\frac{e}{2\pi}\frac{x_i}{r^2} \era \eeq and the total
magnetic flux attached to $N$ charged particles is \beq \Phi=\int_V
d^2 x B(x)=e\te N. \eeq We note that the both $L_{CS}$ (the
lagrangian in Chern-Simons theory) and $L_\te$ lead to fractional
statistics by the same mechanism of attaching a magnetic flux to the
charged particles but the physics they describe is quite different.
We remark, for example, that in this theory the interaction
potential is an object of considerable interest \cite{GC}. The
potential has confining nature; i. e., it grows to infinity when the
natural separation of the physical degrees of freedom grow, but in
the Maxwell-Chern-Simons theory, the Chern-Simons term turns the
electric and magnetic fields massive leading to a screening
potential between static charges.
\section{Anyons and Fuzzy Two-Sphere}
\hspace{.3in}Now, let us consider exotic particle moving on a
two-sphere instead of a plane in the background of a monopole put at
the origin. First, the two-sphere is $S^2\sim CP^1
=\frac{SU(2)}{U(1)}$ and the representations of $SU(2)$ are given by
the standard angular momentum theory.

The coordinates of fuzzy two-sphere are given by the $SU(2)$ algebra
\beq \bra{rl} \lbrack X_i ,X_j \rbrack =i\al\ep_{ijk}X_k ,&X_i =\al
L_i , \era \eeq $L_i$ is the total angular momentum with the
representation to be the spin $\ell$ and $\al$ is a dimensionful
constant. We note that around the north pole of $S^2$ labeled by
$L_3 =\ell$, the fuzzy two-sphere algebra becomes a noncommutative
plane if $\ell\rightarrow\infty$, \beq \lbrack X_i ,X_j \rbrack
=i\al^2 \ell\ep_{ij}I, \eeq with $I$ is the identity.
\subsection{Connection}
\hspace{.3in}To construct the connection which goes to the
generalized connection given above when the radius of fuzzy
two-sphere goes to infinity we use the first Hopf map as known in
the literature which is a map from $S^3$ to $S^2$ and naturally
introduces a $U(1)$ bundle on $S^2$. Then, the two-sphere can be
parameterized by two complex coordinates $u_\al$ such that
$u_\al^\star u_\al =1$ with $u_\al \sim e^{i\te}u_\al$. A spatial
coordinate $x_i$ on $S^2$ with radius $r$ is written in terms of
$u_\al$'s as \beq x_i =ru^\dagger \sigma_i u \eeq with $\sigma$ are
Pauli matrices. The vector potential on $S^2$ is \beq A_i dx_i
=-i\ga u_\al^\star du_\al , \eeq with $\ga$ is integer due to the
Dirac quantization rule and
$u^\star_\al$ is the complex conjugate of $u_\al$.\\

Thus, the Hopf spinor satisfying (8) is given by \beq u=\Big(
\bra{ll}
u_1\\
u_2 \era\Big)=\frac{1}{\sqrt{2r(r+x_3)}}\Big( \bra{ll}
r+x_3\\
x_1+ix_2 \era\Big)e^{i\chi}, \eeq $e^{i\chi}$ is a $U(1)$ phase. The
connection is defined as \beq A_i dx_i =-i\frac{\hbar}{e}
u_\al^\star du_\al =\frac{\hbar}{2er(r+x_3)}\ep_{ij3}x_j dx_i . \eeq
By leading with the motion of an exotic particle (charged
particle-magnetic flux composite) on two-sphere, the monopole charge
is $\frac{\hbar}{2e}=\frac{\te}{4\pi}$ which is identified with the
connection in two dimensional space for $r\rightarrow\infty$
discussed in section 2. By generalizing the spinor to
$(2S+1)$-components spinor $u_{(S)}$, the monopole charge becomes
$\frac{\te}{4\pi}=\frac{\hbar S}{e}$ and $x_i
=\frac{1}{S}ru_{(S)}^\dagger \sigma_{(S)_i} u_{(S)}$, $x_i x_i
=r^2$, where $\sigma_{(S)_i}$ is the spin $S$ representation of
$SU(2)$.

Now, for simplicity we consider a static particle at $\bf{x'}$. The
magnetic and electric fields are given in (4) and the
charge-magnetic dipole is defined by the current \beq\bra{lr} J_0
=e\delta^{(3)}(\bf{x}-\bf{x'})& J_i
=\frac{\phi}{e}\ep_{im}\partial_m J_0 , \era\eeq $\phi$ is the
dipole's moment.
\subsection{Generalized Maxwell Theory}
\hspace{.3in}The Hamiltonian of this system is written as follows
\beq H=\frac{1}{2mr^2 }M_i M_i +\int d^3 x
(-\frac{1}{2}F_{i0}F^{i0}+\frac{1}{4}F_{ij}F^{ij}-\frac{\te}{2}\ep_{ij}J^{0}F^{ij})
\eeq such that for a static point like particle $J_i =0$ and the
primary constraint is $$\pi^0 =0$$ which leads to the secondary
constraint $$\partial_i \pi^i -J^0 =0$$ with $\pi^\mu$ is the
canonical momentum of gauge field $A^\mu$. $M_i$ is the orbital
angular momentum of the charged particle \beq\bra{ll}
M_i &=\ep_{ijk}x_j (-i\hbar\partial_k +eA^\te _k)\\ \\
&=\ep_{ijk}x_j (-i\hbar\partial_k +eA _k
+\frac{e\te}{2}\ep_{knm}F^{nm}+\frac{e\te}{2}\ep_{kn0}F^{n0})
\era\eeq where $i,j,k,n,m=1,2,3$ and $A^\te _k$ is the generalized
connection. The strength field $F^{\mu\nu}$ is \beq\bra{ll}
F^{nm}&=-\frac{\te}{4\pi}\ep_{nml}\frac{x_l}{r^3}\\ \\
F^{n0}&=\frac{\te}{4\pi r(r+x_3)} \ep_{nl3}(\dot{x_l}-\dot{r}x_l
\frac{2r+x_3}{r(r+x_3)}), \era\eeq we note that $\ep_{kn0}F^{n0}=0$
since $\ep_{nl3}\ep_{kn0}=0$ because $l\ne 0$. Then \beq M_i
=\ep_{ijk}x_j (-i\hbar\partial_k +eA_k
-\frac{e\te^2}{4\pi}\frac{x_k}{r^3}). \eeq Thus the Hamiltonian (13)
of this system is reduced to \beq H=\frac{1}{2mr^2 }M_i M_i
+\frac{1}{2}\int d^3 x (E_i ^2 +B^2 ), \eeq with $\ep_{oij}J^0
F^{ij} = \ep_{oij}J^0 \frac{-\te}{4\pi}\ep^{ijk}\frac{x_k}{r^3}
=2\frac{-\te}{4\pi}\delta_{0k}J^0\frac{x_k}{r^3} =0$ in (13) since
$k=1,2,3\ne 0$. Accordingly to (4,5), we calculate the second term
of $H$ in three-dimensional space and the Hamiltonian is \beq H =H_0
+e^2 \te^2 N +\frac{e^2 r}{3\pi},\eeq with \beq H_0 =\frac{1}{2mr^2
}M_i M_i .\eeq The remark we get from this subsection is that the
Hamiltonian is different from the one describing QHE and they are
identified ($H\sim H_0$) only if $N,r\longrightarrow 0$.

\subsection{Realization of Fuzzy Two-Sphere}
\hspace{.3in}First, we remark that the orbital angular momentum of
the particle $M_i $ given by (14) satisfy the following deformed
commutation relations \beq \lbrack M_i , M_j \rbrack=
i\hbar\ep_{ijk}(M_k +\frac{e\te}{2\pi r}x_k ). \eeq This means that
the total angular momentum generalizing the $SU(2)$ algebra should
be defined as \beq L_i =M_i -\frac{e\te}{2\pi r}x_i \eeq and we get
\beq\bra{lll}
\lbrack L_i , L_j \rbrack = i\hbar\ep_{ijk}L_k \\ \\
\lbrack L_i , M_j \rbrack = i\hbar\ep_{ijk}M_k \\ \\
\lbrack L_i , x_j \rbrack = i\hbar\ep_{ijk}x_k . \era\eeq
Consequently, by simple calculation we find that
$$\lbrack L_i , H\rbrack =0,$$ then $SU(2)$ symmetry is generated
by $L_i$. We also see that \beq\bra{lll}
M_i M_i &=L_i L_i - (\frac{e\te}{2\pi})^2 \\\\
&=\hbar^2 (l(l+1)-4S^2), \era\eeq with $\ell$ is the eigenvalue of
$L_3$. In what follows we suggest that $\ell=n+2S$ with
$n=0,1,2...$.

The noncommutative Geometry known as fuzzy two-sphere is described
by the guiding center coordinates since the exotic particle obeys
the cyclotron motion as well-known in the planar system. These
coordinates are defined as \beq X_i =\frac{2\pi r}{e\te}L_i , \eeq
then they are related to the commutative coordinates by \beq X_i
=\frac{2\pi r}{e\te}M_i -x_i . \eeq They satisfy the following
commutative relations \beq \lbrack X_i , X_j \rbrack
=i\hbar\ep_{ijk}\frac{2\pi r}{e\te}X_k , \eeq and the fuzzy
two-sphere is satisfied for the motion of exotic particle on
two-sphere. Its radius is given by the quadratic Casimir of $SU(2)$
\beq r^2 =\hbar^2 (\frac{2\pi r}{e\te})^2 2S(2S+1). \eeq According
to (23,24), we get the radius of the cyclotron motion in the $n$-th
level \beq r^c _n =\frac{2\pi r}{e\te}\hbar\sqrt{2S(2n+1)+n(n+1)}.
\eeq For the lowest level we get \beq r^c _0 =\frac{r}{\sqrt{2S}},
\eeq which is identified with the one obtained in the lowest Landau
Level discussed in \cite{fuzzyQHE}. Also we remark that $r^c_0$ is
much smaller that $r$ in the strong magnetic field limit, and $x_i$
are identified
with $X_i$.\\

\subsection{Energy in Fuzzy Two-Sphere Case}
\hspace{.3in}Owing to (23), the energy eigenvalue of $H$ (18) is
\beq E_n =\frac{\hbar^2 }{2mr^2 }(2S(2n+1)+n(n+1))+e^2 \te^2 N
+\frac{e^2 r}{3\pi}. \eeq Then we notice that this model could be
identified with the one treated in the references \cite{fuzzyQHE}
only in the following case: If both of the radius of fuzzy
two-sphere and the number of charges $N$ are too small; i.e.
$r,N\longrightarrow 0$. Also, we note that $n$ in (30) indicates the
level index which could be identified with the Landau level index in
the Chern-Simons theory only for small $r$ and $N$. The variance
energy between the lowest level $n=0$ and the first level is
$$\Delta E=\frac{\hbar^2 (2S+1)}{mr^2}.$$

As remark, the lowest level is also realized in our system which is
identified to the Lowest Landau level phenomena if the number of
charges is too small and $r\longrightarrow 0$ with $\frac{S}{mr}\gg
1$ and the energy induced by the dynamic gauge field is ignored.
Otherwise, if the above case is not satisfied; i.e. $r\gg 1$ or
$N\gg 1$, the model is now totally different. Thus the variance
energy of the system is $$\Delta E=0,$$ and the energy is dominated
by the one of the two last terms of (28) or both. Then the energy is
\beq E_n =e^2 \te^2 N +\frac{e^2 r}{3\pi},\phantom{~~~~}\forall n.
\eeq We notice here that the variance energy of the system will
depend only on the variance of the number of particles or the
radius.

Consequently, dealing with the case of exotic particles system in
which we introduce a generalized connection put at the origin of
two-sphere we get a noncommutative geometry. The energy obtained in
this model is very special and too different from the one obtained
in QHE case since the gauge field in this system is dynamic. We
notice that the energy of gauge field dominates when the radius of
fuzzy two-sphere goes to infinity; i.e. the flat D2-brane which is a
dual of fuzzy two-sphere has high energy. We remark that this result
is definitely different from the one could be obtained in the case
of QHE. In this latter case if $r\longrightarrow\infty$ the fuzzy
two-sphere goes to flat D2-brane having low energy which goes to
zero.
\subsection{Potential}
\hspace{.3in}We complete this section by giving another interesting
remark. As known, the potential has confining nature in
two-dimensional space when the generalized connection is introduced
instead of adding CS-term; i. e., the potential grows to infinity
when the natural separation of the physical degrees of freedom
grows.

After giving the energy we may now proceed to discuss the
interaction energy between pointlike sources in the model under
consideration. This can be done by computing the expectation value
of the energy operator H in a physical state $\mid\Om\rangle$ by
following the mechanism used in \cite{pot}. We consider the stringy
gauge-invariant $\mid\bar{\Psi}(y) \Psi(r)\rangle$ state, \beq
\mid\Om\rangle\equiv \mid\bar{\Psi}(y) \Psi(y')\rangle
=\mid\bar{\psi}(y)e^{-ie\int\limits_{y}^{y'}dz^i A_i (z)}
\psi(y')\mid 0\rangle, \eeq where $\mid 0\rangle$ is the physical
vacuum state and the integral is to be over the linear spacelike
path starting at $y$ and ending at $y'$, on a fixed time slice. Note
that the strings between exotic particles have been introduced to
have a gauge-invariant state $\mid\Om\rangle$, in other terms, this
means that the elementary particles (bosons or fermions) are now
dressed by a cloud of gauge fields.

From the foregoing Hamiltonian discussion, we first note that \beq
\pi_i \mid\bar{\Psi}(y) \Psi(y')\rangle=\bar{\Psi}(y) \Psi(y') \pi_i
\mid 0\rangle + e\int\limits_{y}^{y'} dz_i \delta^3
(x-z)\mid\bar{\Psi}(y) \Psi(y')\rangle . \eeq Owing to (17,31) and
the fact that we consider a static pointlike particle; so $\pi_i
=F_{0i}=E_i$, we get the expectation value of the Hamiltonian as
\beq \langle\Om\mid H\mid\Om\rangle=\langle 0\mid H\mid 0\rangle
+\frac{e^2}{2}\int d^3 x \Big( \int\limits_{y}^{y'} dz_i \delta^3
(x-z)\Big)^2 , \eeq with $x$ and $z$ are three-dimensional vectors.
Remembering that the integrals over $z_i$ are zero except on the
contour of integrations.

The last term of (34) is nothing but the Coulomb interaction plus an
infinite self-energy term. In order to carry out this calculation we
write the path as $z = y+\al(y -y')$ where $\al$ is the parameter
describing the contour. By using the spherical coordinates the
integral under square becomes \beq \int\limits_{y}^{y'} dz_i
\delta^3 (x-z)=\frac{y-y'}{|y-y'|^2}\int\limits_{0}^{1} d\al
\frac{1}{\al}\de(|y-x|,\al|y'-y|)\sum\limits_{\ell,m}Y^* _{\ell
m}(\te' ,\phi')Y _{\ell m}(\te ,\phi). \eeq Using the usual
properties for the spherical harmonics and after subtracting the
self-energy term, we obtain the potential as \beq V=-\frac{e^2
}{4\pi} \frac{1}{|y' -y|} \eeq

This result lets us to draw attention to the fact that with fuzzy
two-sphere the generalized Maxwell theory doesn't have confining
nature any more which was a special property for anyons described by
generalized Maxwell theory in two-dimensional space. Thus the
problem of confinement could be solved by considering the two-sphere
in stead of two-dimensions.
\section{Conclusion}
\hspace{.3in}In this paper, we have used the generalized Maxwell
theory on two-sphere instead of two-dimensional space. This leaded
to get some results totally different from those gotten in the case
of QHE in high dimensions \cite{fuzzyQHE}. By considering the exotic
particles described by generalized Maxwell theory, the energy
produced by the gauge field is involved in the energy of the system
(18) depending on the number of charges and the radius of the
sphere. We remark that the energy of gauge field dominates when the
radius of two-sphere goes to infinity; i. e. the energy of flat
D2-brane is generated by the gauge field leading to high energy. We
also notice that the energy becomes more higher if the number of
charges is large. Another important remark is that with fuzzy
two-sphere the static potential for two opposite charged exotic
particles loses automatically its confining nature without adding
the CS-term; i. e. by plunging the generalized Maxwell theory in
high dimensions the potential has screening nature.\\
\section*{Acknowledgments}
The author would like to thank Robert De Mello Koch for interesting
discussions and his hospitality at WITS university where a part of
this work was done.

\end{document}